%% file: PionMAPTMD22.tex
\documentclass[aps,reprint,nofootinbib,superscriptaddress]{revtex4}

\include{settings}

\allowdisplaybreaks[2]

\begin{document}

\title{Extraction of Pion Transverse Momentum Distributions from Drell-Yan data
  \\ \vspace{0.2cm}
\normalsize{\textmd{The \textbf{MAP} Collaboration}\footnote{The MAP acronym stands for ``Multi-dimensional Analyses of Partonic distributions''. It refers to a collaboration aimed at studying the three-dimensional structure of hadrons. The public codes released by the collaboration are available at \href{https://github.com/MapCollaboration}{https://github.com/MapCollaboration}.}}
}

\author{Matteo Cerutti}
\thanks{Electronic address: matteo.cerutti@pv.infn.it -- \href{https://orcid.org/0000-0001-7238-5657}{ORCID: 0000-0001-7238-5657}}
\affiliation{Dipartimento di Fisica, Universit\`a di Pavia, via Bassi 6, I-27100 Pavia, Italy}
\affiliation{INFN - Sezione di Pavia, via Bassi 6, I-27100 Pavia, Italy}

\author{Lorenzo Rossi}
\thanks{Electronic address: lorenzo.rossi@pv.infn.it -- \href{https://orcid.org/
0000-0002-8326-3118}{ORCID: 0000-0002-8326-3118}}
\affiliation{Dipartimento di Fisica, Universit\`a di Pavia, via Bassi 6, I-27100 Pavia, Italy}
\affiliation{INFN - Sezione di Pavia, via Bassi 6, I-27100 Pavia, Italy}

\author{Simone Venturini}
\thanks{Electronic address: simone.venturini01@universitadipavia.it -- \href{https://orcid.org/
0000-0002-4105-7930}{ORCID: 0000-0002-4105-7930}}
\affiliation{Dipartimento di Fisica, Universit\`a di Pavia, via Bassi 6, I-27100 Pavia, Italy}
\affiliation{INFN - Sezione di Pavia, via Bassi 6, I-27100 Pavia, Italy}

\author{Alessandro Bacchetta}
\thanks{Electronic address: alessandro.bacchetta@unipv.it -- \href{https://orcid.org/0000-0002-8824-8355}{ORCID: 0000-0002-8824-8355}}
\affiliation{Dipartimento di Fisica, Universit\`a di Pavia, via Bassi 6, I-27100 Pavia, Italy}
\affiliation{INFN - Sezione di Pavia, via Bassi 6, I-27100 Pavia, Italy}

\author{Valerio Bertone}
\thanks{Electronic address: valerio.bertone@cea.fr -- \href{https://orcid.org/0000-0003-0148-0272}{ORCID: 0000-0003-0148-0272}}
\affiliation{IRFU, CEA, Universit\'e Paris-Saclay, F-91191 Gif-sur-Yvette, France}

\author{Chiara Bissolotti}
\thanks{Electronic address: cbissolotti@anl.gov --
  \href{https://orcid.org/0000-0003-3061-0144}{ORCID: 0000-0003-3061-0144}}
\affiliation{HEP Division, Argonne National Laboratory, 9700 S. Cass Avenue, Lemont, IL, 60439 USA}

\author{Marco Radici}
\thanks{Electronic address: marco.radici@pv.infn.it -- \href{https://orcid.org/0000-0002-4542-9797}{ORCID: 0000-0002-4542-9797}}
\affiliation{INFN - Sezione di Pavia, via Bassi 6, I-27100 Pavia, Italy}

\begin{abstract}
  We map the distribution of unpolarized quarks inside a unpolarized pion
  as a function of the quark's transverse momentum, encoded in unpolarized
  Transverse Momentum Distributions (TMDs). We extract the pion TMDs from
  available data of unpolarized pion-nucleus
Drell--Yan processes, where the cross section is differential in the lepton-pair transverse momentum.
In the cross section, pion TMDs are convoluted with nucleon TMDs that we consistently take from our previous studies.
We obtain a fairly good agreement with data.
We present also predictions for pion-nucleus scattering that is being measured by the COMPASS Collaboration.
\end{abstract}

\maketitle
\section{Introduction}
\label{s:intro}

The pion is the simplest of all hadrons and together with the nucleon constitutes one of the most fundamental entities in the visible Universe. In the Standard Model, both particles are built as bound states of the quark and gluon degrees of freedom of Quantum ChromoDynamics (QCD). However, in this context the pion plays a unique role since it is the Goldstone boson of chiral symmetry breaking. Therefore, it is extremely important to investigate its internal structure and understand how the latter is responsible of the macroscopic differences between the bound state of a pion and of a nucleon (see, e.g., Ref.~\cite{Horn:2016rip} for a review).


The internal structure of the pion can be described in terms of Parton
Distribution Functions (PDFs). Starting from the 1990s,
PDFs have been extracted from data in various
papers~\cite{Gluck:1991ey,Sutton:1991ay,Gluck:1997ww,Gluck:1999xe,
Wijesooriya:2005ir,Aicher:2010cb,Barry:2018ort,
Novikov:2020snp,Cao:2021aci,Barry:2021osv,Bourrely:2022mjf}. In spite of this
extensive literature, the structure of the pion is known to a much less extent
than that of the proton, due to the scarcity of data on high-energy scattering
processes involving pions.

While
PDFs describe the distribution of quarks and gluons as a function of only their momentum component longitudinal to the parent hadron,
Transverse Momentum Distributions (TMDs) describe
the distribution in three-dimensional momentum space. If the knowledge of the
one-dimensional structure of the pion is limited, almost nothing is known
about its three-dimensional structure.
Model calculations of
pion TMDs have been discussed in
Refs.~\cite{Pasquini:2014ppa,Noguera:2015iia,Lorce:2016ugb,Bacchetta:2017vzh,Ceccopieri:2018nop,Ahmady:2019yvo,Kaur:2019jfa,Shi:2020pqe}.

The extraction of TMDs from
data is based on TMD factorization theorems and is more challenging than that
of collinear PDFs.
For proton TMDs, factorization theorems have been proven for Semi-Inclusive Deep-Inelastic Scattering (SIDIS), for Drell-Yan (DY) lepton-pair production in hadronic collisions, and for semi-inclusive electron-positron annihilations (see Ref.~\cite{Collins:2011zzd} and references therein). Recently, very accurate studies of proton unpolarized TMDs have been released~\cite{Bacchetta:2017gcc,Scimemi:2017etj,Bertone:2019nxa,Scimemi:2019cmh,Bacchetta:2019sam,Bury:2022czx,Bacchetta:2022awv}, some of which are based on a global analysis of such processes.
For pion TMDs, data are available only for the DY process and only two
analyses have been published~\cite{Wang:2017zym,Vladimirov:2019bfa}.


In this paper, we present an extraction of unpolarized pion quark TMDs by analyzing for the first time the whole set of available data for the DY lepton-pair production in $\pi^-$-nucleus collisions, obtained from the
E615~\cite{Stirling:1993gc} and E537~\cite{Anassontzis:1987hk} experiments.
The cross section differential in the lepton-pair transverse momentum can be written as
a convolution of a unpolarized proton TMD and a unpolarized pion TMD. For
the proton TMD, we use the result
recently obtained by the MAP
collaboration~\cite{Bacchetta:2022awv} and we extract the pion TMD from data using the
same formalism.
With respect to Ref.~\cite{Vladimirov:2019bfa}, we use more data and different
prescriptions for the implementation of TMD evolution. With respect to
Ref.~\cite{Wang:2017zym}, we use more data, a higher theoretical accuracy, a up-to-date
extraction of the proton TMDs, and we consistently use the same Collins--Soper
kernel for proton and pion TMDs.

\section{Formalism}
\label{s:formalism}


In the DY process
\begin{equation}
\label{e:DYproc}
  h_A(P_A) + h_B(P_B) \rightarrow \gamma^*(q) + X \rightarrow \ell^+(l) + \ell^-(l') + X
\end{equation}
the collision between two hadrons $h_A$ and $h_B$ with four-momenta $P_A$ and $P_B$, respectively, and center-of-mass energy squared $s=(P_A + P_B)^2$, produces a neutral vector boson $ \gamma^*/Z$ with four-momentum $q$ and large invariant mass $Q = \sqrt{q^2}$. The vector boson eventually decays into a lepton and an antilepton with four-momenta constrained by momentum conservation, $q = l + l'$.  The cross section of this process can be written in terms of two structure functions  $F^1_{UU}$,  $F^2_{UU}$. Being $M$ the mass of the incoming hadrons and $\qT$ the transverse component of the vector boson momentum, in the kinematic region where $M^2 \gg Q^2$ and $\qT^2 \ll Q^2$  the structure function $F^2_{UU}$ is suppressed. Therefore,  the cross section can be expressed as
\begin{equation}
\label{e:DYxs}
\begin{split}
 \frac{d\sigma^{DY}}{d|\qT|dydQ} & \simeq \frac{16\pi^2\alpha^2}{9Q^3} |\qT| F^1_{UU}(x_A,x_B,\qT,Q)
 \\ & =
\frac{16\pi^2\alpha^2}{9Q^3}|\qT| \frac{x_A x_B}{2\pi} \mathcal{H}^{DY}(Q;\mu) \sum_a c_a(Q^2)  \int d |\bT| |\bT| J_0(|\qT| |\bT|) \hat{f}_{1}^a(x_A,\bT ^2;\mu,\zeta_A) \hat{f}_1^{\bar{a}}(x_B,\bT ^2;\mu,\zeta_B),
\end{split}
\end{equation}
where $\alpha$ is the electromagnetic coupling, $y$ is the pseudorapidity of the vector boson, $x_{A,B} = \frac{Q}{\sqrt{s}} e^{\pm y}$, $\mathcal{H}^{DY}$ is the hard factor and $\hat{f}_1^a$ is the Fourier transform of the unpolarized TMD PDF for flavor $a$,\footnote{For the definition of the Fourier transform see Ref.~\cite{Bacchetta:2022awv}.} which depends on the renormalization and rapidity scales $\mu$ and $\zeta$, respectively. The summation over $a$ in Eq.~\eqref{e:DYxs} runs over the active quarks and antiquarks at the hard scale $Q$, with $c_a (Q^2)$ the corresponding electroweak charges~\cite{Bacchetta:2022awv}.




The dependence of $\hat{f}_1$ on the scales $\mu$ and $\zeta$ arises from the removal of the ultraviolet and rapidity divergences in its operator definition. In the $\overline{\mathrm{MS}}$ renormalization scheme, it turns out that the choice of the initial scale $\mu_0 = \sqrt{\zeta_0} = \mu_b (|\bT|) = 2 e^{-\gamma_E}/|\bT|$ (with $\gamma_E$ the Euler constant) greatly simplifies the expression of $\hat{f}_1$~\cite{Bacchetta:2019sam}. However, in order to prevent $\mu_b$ from becoming larger than $Q$ at small $|\bT|$ and/or hitting the Landau pole at large $|\bT|$ it is necessary to introduce an {\it ad-hoc} prescription $b_* (\bT^2)$ to avoid these limits. Then, the TMD PDF can be simply rewritten as~\cite{Bacchetta:2019sam}
\begin{eqnarray}
\hat{f}_{1} (x,\bT ^2;\mu,\zeta) &= &\left[ \frac{\hat{f}_{1} (x,\bT ^2;\mu,\zeta)}{\hat{f}_{1} (x,b_* (\bT^2);\mu,\zeta)} \right] \, \hat{f}_{1} (x,b_* (\bT^2);\mu,\zeta) \nonumber \\
&\equiv &f_{1NP} (x, \bT^2; \zeta) \, \hat{f}_{1} (x,b_* (\bT^2);\mu,\zeta) \; ,
\label{e:TMDb*}
\end{eqnarray}
which effectively defines the nonperturbative part $f_{1NP}$ of the TMD PDF.

For the collision of a pion and a nucleus (that in first approximation is described as a collection of free nucleons), the cross section of Eq.~\eqref{e:DYxs} involves the TMD PDFs $\hat{f}_{1p}^a$ and $\hat{f}_{1\pi}^a$ of a quark $a$ in the proton and in the pion, respectively.
As for the proton, we use for $\hat{f}_{1p}^a$ the recent global extraction of the MAP Collaboration~\cite{Bacchetta:2022awv} at next-to-next-to-next-leading-logarithm (N$^3$LL) accuracy, adopting the same $b_*(\bT^2)$ prescription and the same parametrization of the nonperturbative part $f_{1NP}^p$.
As for the pion, for $\hat{f}_{1\pi}^a$ we consistently retain the same $b_*(\bT^2)$ prescription and the same nonperturbative component $g_K (\bT^2)$ of the Collins--Soper evolution kernel as in Ref.~\cite{Bacchetta:2022awv}, but we use the following expression for $f_{1NP}^{\pi}$:
\begin{equation}
\label{e:pif1NP}
f_{1NP}^{\pi}(x,\bT^2; \zeta) = e^{-g_{1\pi}(x) \frac{\bT^2}{4}} \bigg{[} \frac{\zeta}{Q_0}\bigg{]}^{g_K (\bT^2)/2} = e^{-g_{1\pi}(x) \frac{\bT^2}{4}} \bigg{[} \frac{\zeta}{Q_0}\bigg{]}^{-g_2^2 \frac{\bT^2}{4}} \; ,
\end{equation}
where $Q_0$ is an arbitrary starting scale that we choose to be 1 GeV, and $g_2$ is a parameter taken from Ref.~\cite{Bacchetta:2022awv}, as mentioned above. The $x$-dependence of the width $g_{1\pi}$ is given by
\begin{equation}
\label{e:Gausswidth}
g_{1\pi}(x) = N_{1 \pi} \frac{x^{\sigma_{\pi}} (1-x)^{\alpha^2_{\pi}}}{\hat{x}^{\sigma_{\pi}} (1-\hat{x})^{\alpha^2_{\pi}}} \; ,
\end{equation}
with $\hat{x}=0.1$.
In conclusion, we have a total of 3 free parameters to fit to data: $N_{1 \pi}$, $\sigma_{\pi}$ and $\alpha_{\pi}$.

\section{Data selection}
\label{s:data}

\begin{figure}[H]
	\centering
	\includegraphics[width=0.44\textwidth]{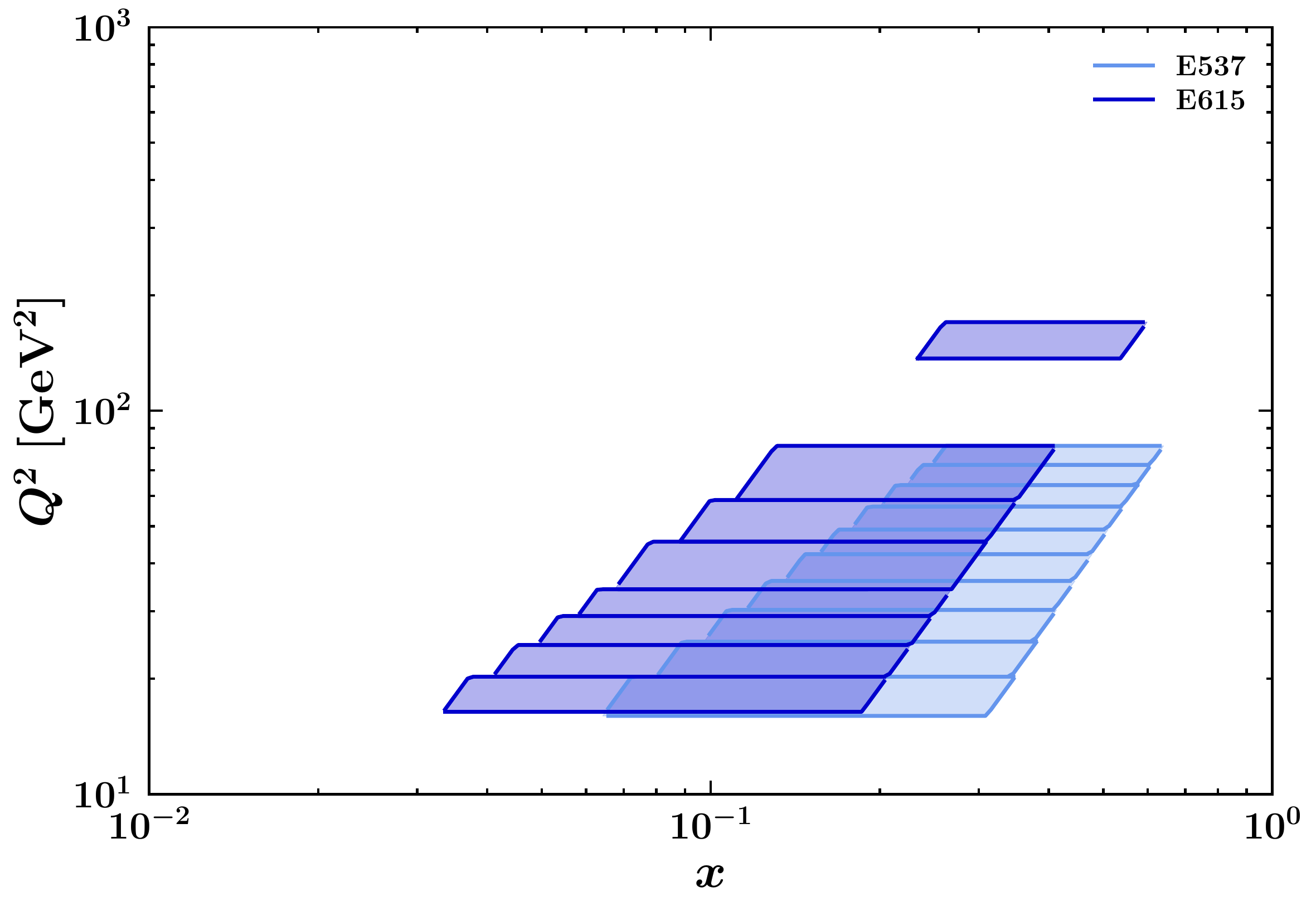}
	\caption{The coverage of  E615~\cite{Stirling:1993gc} (blue) and E537~\cite{Anassontzis:1987hk} (light blue) data in the plane $x$-$Q^2$.}
	\label{f:Data_Coverage}
\end{figure}

In this Section we describe the experimental data included in our
analysis. We consider data from the two DY experiments
(E615~\cite{Stirling:1993gc} and E537~\cite{Anassontzis:1987hk}) that used
$\pi^-$ beams impinging on tungsten targets. The coverage of these data in the
$x$-$Q^2$ plane is shown in Fig.~\ref{f:Data_Coverage}.

Since our study is based on TMD factorization, which is applicable only in the kinematical region  $|\bm{q}_T| \ll Q$, we apply to the data sets the following cut:
\begin{equation}
	\label{e:Cut}
	\frac{|\bm{q}_T|}{Q} < 0.3 + \frac{0.6}{Q} \; .
\end{equation}
This cut is slightly different from the analogous one used in the proton TMD
extraction: since the availability of data is limited, this cut is a
good compromise between the necessity of including more data and the necessity
of staying within the limits of applicability of TMD factorization.

In order to avoid the kinematical region of invariant masses around the $\Upsilon$ resonance, we also exclude those bins of the DY experiment E615 for which $9.00 \text{~GeV} < Q < 11.70 \text{~GeV}$. The numbers of experimental data before and after the application of these cuts are reported in Table~\ref{t:Table_Data}, together with other useful details regarding the experimental processes, like the definition of the observables and the values of $\sqrt{s}$ and $Q$.
\renewcommand{\tabcolsep}{6pt}
\renewcommand{\arraystretch}{1.5}
\begin{table}[H]
	\begin{center}
		\begin{tabular}{c||c|c|c|c|c|c|c}
			\hline
			Experiment & $N_{\mathrm{dat}}$ & $N_{\mathrm{cut}}$  &Observable & $\sqrt{s}$ [GeV] & $Q$ range [GeV] & $N_{Q\mathrm{-bin}}$ & $x_F$ range  \\
			\hline \hline
			E615 & 155 & 74 & $d^2 \sigma / dQ d|\qT|$ &
                        21.8 &4.05~-~13.05 & 8 &0.0~-~1.0\\ \hline
			E537 & 150 & 64 & $d^2 \sigma / dQ d \qT^2$ & 15.3 & 4.0~-~9.0 & 10 &-0.1~-~1.0\\
			\hline
		\end{tabular}
		\caption{For each experiment we indicate the original number $N_{\mathrm{dat}}$ of data points, the number $N_{\mathrm{cut}}$ of data points included in the fit after applying the kinematical cuts, the delivered observable, the center-of-mass energy $\sqrt{s}$, the range in invariant mass $Q$, the number of considered bins in $Q$ and the integration range in $x_F$. The width of the $Q$--bins for E537 is $0.5$~GeV, for E615 the width increases with $Q$: for the first four bins it is $0.45$~GeV, for the next two ones it is $0.9$~GeV, and for the last two ones it is  $1.35$~GeV. }
		\label{t:Table_Data}
	\end{center}
\end{table}


Each of the considered data sets has systematic and statistical uncertainties.
The statistical uncertainties are quite large for both E537 (8\%) and E615
(16\%). We choose to treat them as uncorrelated, while we treat the systematic
uncertainties as fully correlated.
For small $|\bT| \ll 1 / \Lambda_{\text{QCD}}$, the TMD PDFs can be matched
onto the related collinear PDFs. Therefore, we must consider the systematic
uncertainties induced by the choice of the collinear PDFs. Consistently with
Ref.~\cite{Bacchetta:2022awv}, for the proton TMD PDF we choose the PDF
parametrization MMHT2014~\cite{Harland-Lang:2014zoa}; for the pion we use the
xFitter20 one~\cite{Novikov:2020snp}. We compute the PDF uncertainties by
using the Hessian method,
and we consider 80\% of them as fully correlated while the remaining 60\% as
uncorrelated, as already done in Ref.~\cite{Bacchetta:2022awv}.

\section{Results}
\label{s:results}

In this section, we present the results for the extraction of the
unpolarized quark TMD PDFs in a pion from a fit of all the existing DY data involving pions
(see Sec.~\ref{s:data}). Since collinear sets for pion PDFs are presently
available only at next-to-leading-order (NLO) accuracy, the TMD PDFs can be
extracted at an accuracy that we defined as N$^3$LL$^-$ (similarly to Ref.~\cite{Bacchetta:2022awv}), i.e., all ingredients are at N$^3$LL accuracy, apart from the evolution of the collinear PDFs.

The error analysis is performed with the so-called bootstrap method, by fitting an ensemble of 200 replicas of experimental data. For consistency with
Ref.~\cite{Bacchetta:2022awv}, we used the proton TMD PDFs extracted in the
fit of Ref.~\cite{Bacchetta:2022awv} and we associate the $i$--th replica of
quark TMD PDFs in the pion
to the same replica in the set of quark TMD PDFs in the proton.

\subsection{Fit quality}
\label{ss:fitqual}

When using the bootstrap method, the full statistical information is contained
in the whole set of 200 replicas of the extracted TMD PDFs. Nevertheless, we
choose as the most representative indicator of the quality of our fit the
$\chi^2$ value of the best fit to the unfluctuated data, $\chi^2_0$ (also
referred as the central replica).

It is possible to decompose the value of $\chi_0^2$ as the sum of two different contributions~\cite{Bacchetta:2022awv},
\begin{eqnarray}
\label{e:chi2terms}
\chi_0^2 = \chi^2_D + \chi^2_{\lambda} \; ,
\end{eqnarray}
where $\chi^2_D$ is the contribution of uncorrelated uncertainties and the penalty $\chi^2_\lambda$ is related to correlated uncertainties.

The breakup of $\chi^2_0$ into its components, normalized to the number of data points surviving the kinematical cuts ($N_\text{cut}$), is reported in Tab.~\ref{t:chi2} for the two experimental sets
included in this analysis.

\begin{table}[h]
\centering
\begin{tabular}{c||c|c|c|c}
\hline
\textbf{Experiments} & $N_{\text{cut}}$ & $\chi_D^2 / N_{\text{cut}}$  &
$\chi_\lambda^2 / N_{\text{cut}}$       & $\chi_0^2 / N_{\text{cut}}$
\\ \hline \hline
E537             & 64 & 1.00      & 0.57 & 1.57 \\ \hline
E615           & 74 & 0.31      & 1.22 & 1.53 \\ \hline
Total & 138 & 0.63 & 0.92 & 1.55 \\ \hline
\end{tabular}
\caption{The break up of the central replica $\chi^2_0$ into components related to uncorrelated ($\chi^2_D$) and correlated ($\chi^2_\lambda$) uncertainties, normalized to the number of data points surviving the kinematical cuts ($N_\text{cut}$).}
\label{t:chi2}
\end{table}

We note that the global $\chi^2_\lambda / N_{\text{cut}} = 0.92$ reported in Tab.~\ref{t:chi2} suggests that the comparison between data and theory is strongly affected by normalization errors. In fact, the partial values of $\chi_D^2$ indicate that the description of the shape of the experimental data is very good (almost perfect for the E615 data set), but there is a systematic disagreement between data and theory in the normalization, which induces high values in the penalty $\chi_{\lambda}^2$.
Since theoretical errors related to the collinear PDFs uncertainty of both pion and proton
are not larger than 5-8$\%$, we think that such a large value of the penalty $\chi_{\lambda}^2$
is given by the correlated systematic uncertainties of the experimental data sets ($\sim 16\%$).
This conclusion is compatible with the findings of Ref.~\cite{Vladimirov:2019bfa}, where the same issue was remarked for the E615 dataset.
Since low-energy fixed-target DY data are properly described in global fits of unpolarized
proton TMDs (see, e.g., Refs.~\cite{Bacchetta:2022awv,Bacchetta:2019sam,Bacchetta:2017gcc,Bertone:2019nxa}),  
this normalization issue does not seem to be related to the fact that E615 and E537 experiments were done at
low invariant masses $Q$.
In Ref.~\cite{Vladimirov:2019bfa}, it was pointed out that the observed issue could be related to
a wrong normalization of the experimental data.

\begin{figure}[h]
\centering
\includegraphics[width=0.8\textwidth]{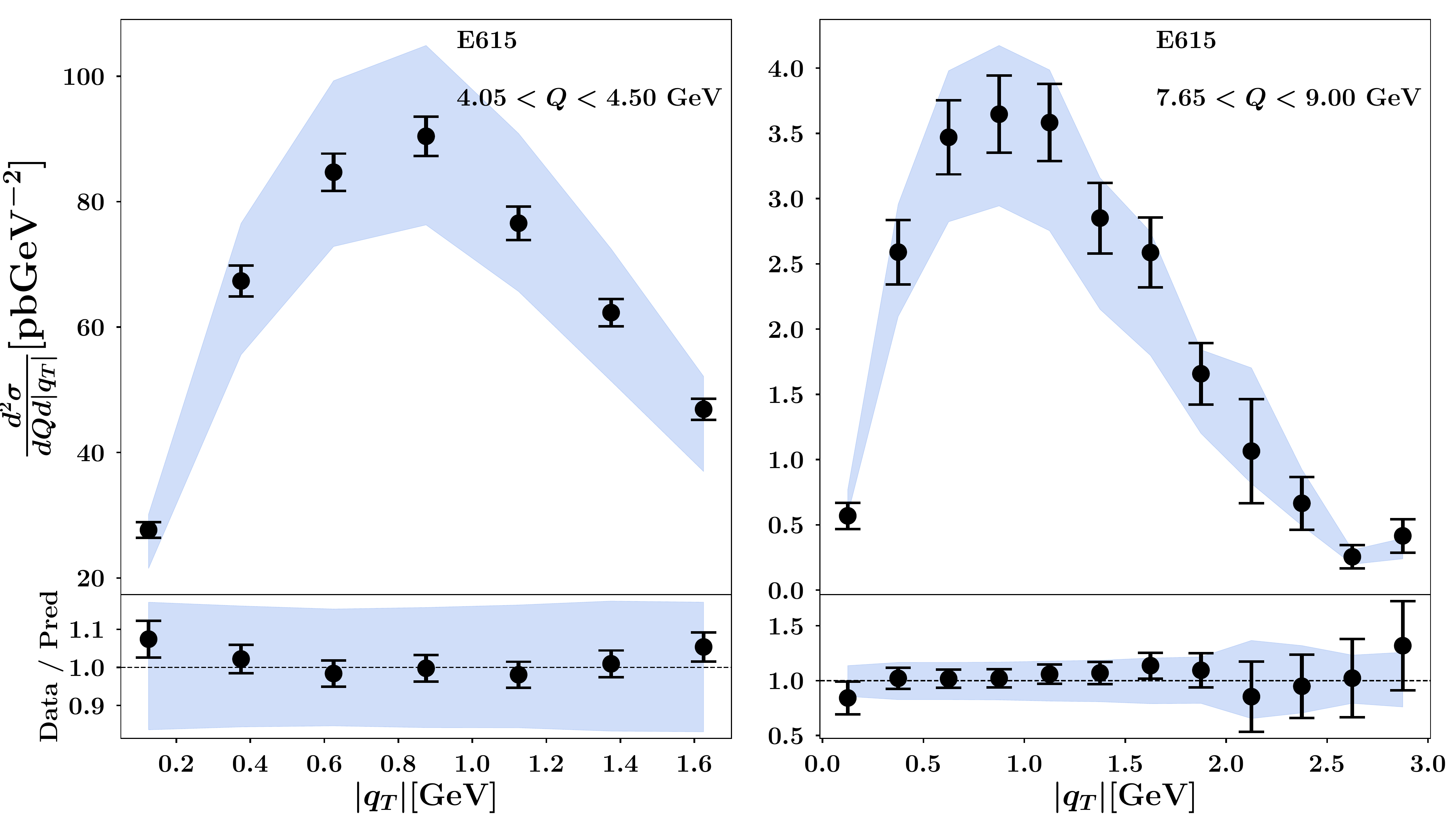}
\caption{Comparison between data (black points) and our fit (colored band) for two different $Q$ bins of the E615 data set. Upper panels: DY cross section differential in $|\qT|$; lower panels: ratio between data and results of the fit. Uncertainty bands correspond to the 68$\%$ CL.}
\label{f:E615_plot}
\end{figure}

In Fig.~\ref{f:E615_plot}, we show the comparison between the result of our fit (colored band) and experimental data (black points) for a selection of $Q$ bins of the E615 data set. In the upper panels, the differential DY cross section is shown as a function of the transverse momentum $|\qT|$ of the virtual vector boson, in the lower panels the ratio between theory and data is displayed. The uncertainty bands
correspond to the 68$\%$ Confidence Level (CL), built by excluding the largest and smallest 16$\%$ of the replicas. As mentioned above, we note that the shape of the experimental data is very well reproduced. The large error band reflects
the large correlated systematic errors of the considered data set.


\begin{figure}[h]
\centering
\includegraphics[width=0.8\textwidth]{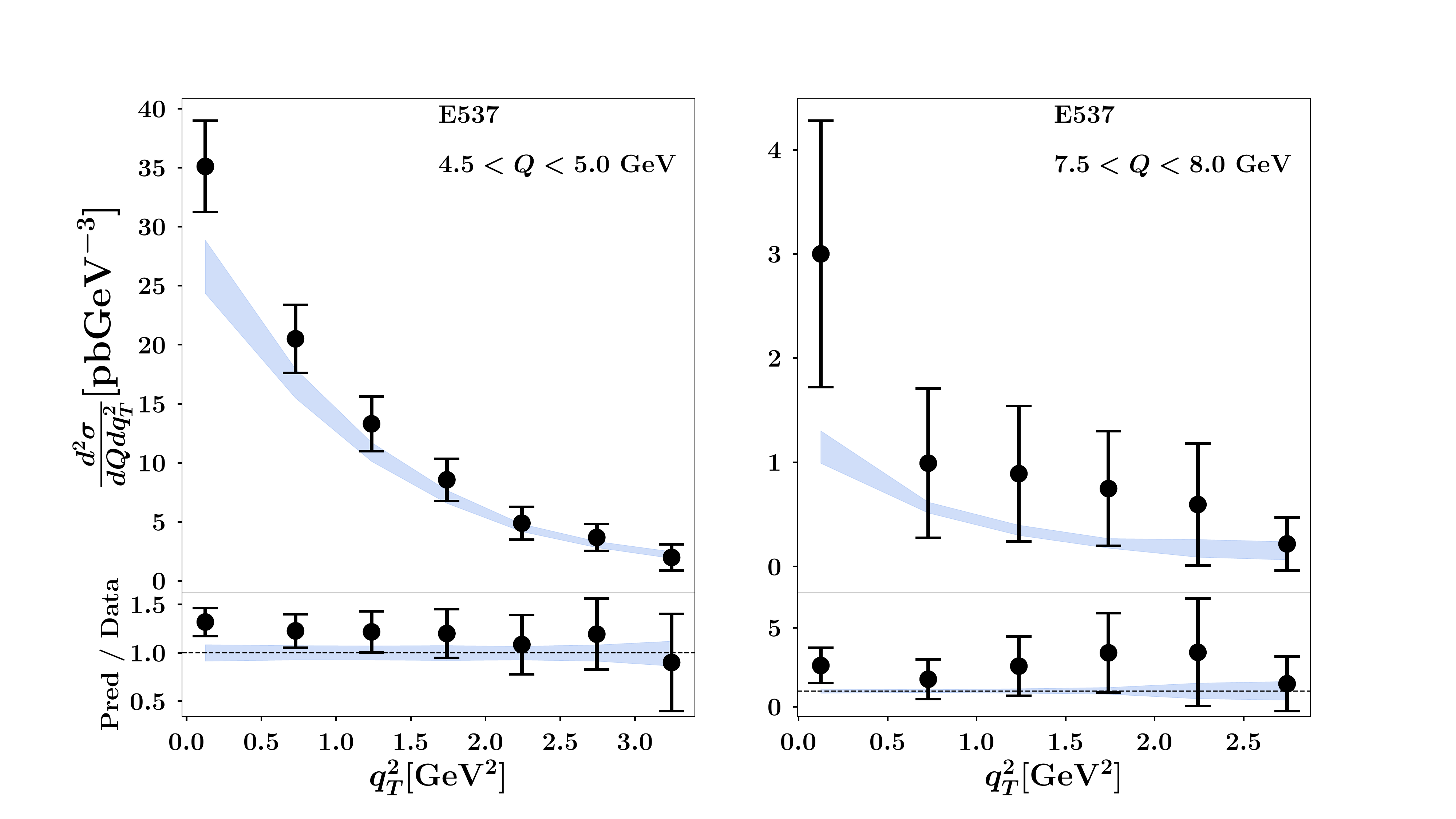}
\caption{Upper panels: comparison between data and theoretical predictions for the DY cross section differential in $\qT^2$ for the \textsc{E537} dataset for different $Q$ bins; uncertainty bands correspond to the 68\% CL. Lower panels: ratio between experimental data and theoretical cross section.}
\label{f:E537_plot}
\end{figure}

In Fig.~\ref{f:E537_plot}, we show the same kind of comparison as in the previous figure but for the E537 data set. Again, in the upper panels we compare the uncertainty bands at 68\% CL from our fit with experimental data (black points) for the DY cross section as function of $\qT^2$ for two different $Q$ bins; the lower panels contain the ratio between fit results and data.

The overall quality of the fit of E537 data is slightly worse than the E615 one. The very similar value of $\chi_0^2 / N_{\text{cut}}$ in this case is the result of different partial contributions. The component $\chi_D^2$ due to uncorrelated uncertainties is three times larger than in the E615 case and gives the largest contribution to $\chi_0^2$. This is reflected in Fig.~\ref{f:E537_plot} where the data points at low $\qT^2$ are not well described by our fit. This discrepancy could be related to the lack of flexibility of our parametrization of the nonperturbative part of the quark TMD PDF in the pion. We explored different models for $f_{1NP}^{\pi}$ in Eq.~\eqref{e:pif1NP}, but with no significant change in the final outcome. On the other side, the contribution of the penalty $\chi_{\lambda}^2$ is less than half of the E615 one. Indeed, the correlated systematic uncertainties of the E537 data set are much smaller and signal that the normalization problem between theory and data is less strong. This is reflected in much smaller uncertainty bands of the fit results, as shown in Fig.~\ref{f:E537_plot}.

We also explored the behaviour of our fit when reducing the accuracy of the theoretical description. As expected, the $\chi_0^2$ worsens (at N$^2$LL, $\chi_0^2/N_{\text{cut}} = 1.72$, at NLL, $\chi_0^2/N_{\text{cut}} = 2.00$) but it shows a nice convergence to the N$^3$LL result in Tab.~\ref{t:chi2}, when reading in reverse order. Moreover, the best values of free parameters remain always within the uncertainty bands, indicating that our fit results are stable.


\subsection{TMD distributions}
\label{ss:TMDs}

We now discuss the quark TMD PDFs in the pion extracted from our fit at
N$^3$LL$^-$. In Tab.~\ref{t:params}, we list the resulting average values and
standard deviations of the three fitting parameters that describe the
arbitrary nonperturbative part of the TMD PDF in Eqs.~\eqref{e:pif1NP}
and~\eqref{e:Gausswidth}. The errors are very large and the parameters are not
very well constrained. As mentioned above, the same quality of results is
obtained by testing different models of nonperturbative
parametrizations. Hence, we conclude that the current set of experimental data is not very sensitive to these degrees of freedom and more data are needed to better constrain them.

\begin{table}[h]
\begin{center}
\begin{tabular}{c|c}
  \hline
  \textbf{Parameter} & \textbf{Average $\pm \bm{\sigma}$ } \\
  \hline \hline
  $N_{1\pi} \ [\text{GeV}^2]$ & 0.47 $\pm$ 0.12 \\
  \hline
  $\sigma_\pi$ & 4.50 $\pm$ 2.25 \\
  \hline
  $\alpha_\pi$ & 4.40 $\pm$ 1.34 \\
  \hline
\end{tabular}
\caption{Average and one standard deviation over the Monte Carlo replicas of the fitting parameters in the nonperturbative part of the quark TMD PDFs in a pion.}
\label{t:params}
\end{center}
\end{table}


In agreement with Ref.~\cite{Vladimirov:2019bfa}, we also find that there are strong correlations among the three different parameters.
This feature points again toward the need for new experimental data to better constrain the quark TMDs in the pion.


\begin{figure}[h]
\centering
\includegraphics[width=1\textwidth]{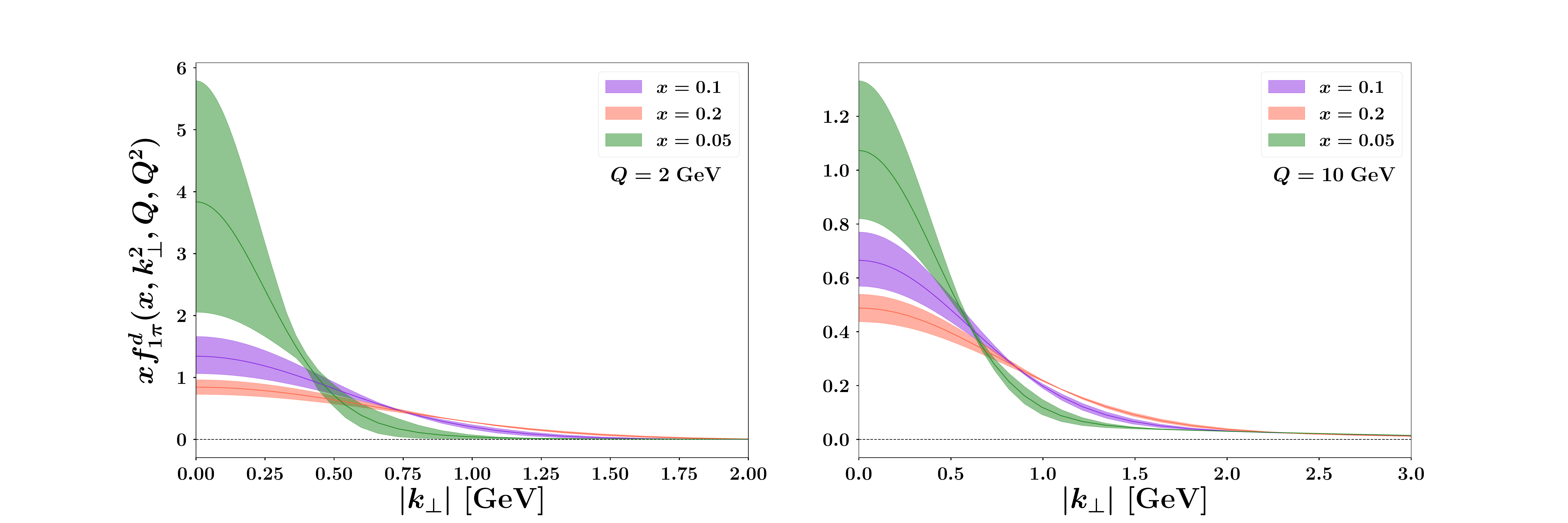}
\caption{The TMD PDF of the down quark in $\pi^-$ at $\mu = \sqrt{\zeta} = Q = 2$ GeV (left panel) and 10 GeV (right panel) as a function of the partonic transverse momentum $|\kperp|$ for $x$ = 0.05, 0.1 and 0.2. The uncertainty bands correspond to the 68\% CL.}
\label{f:tmds}
\end{figure}

In Fig.~\ref{f:tmds}, we show the unpolarized TMD PDF for a $d$ quark in $\pi^-$ at
$\mu = \sqrt{\zeta} = Q = 2$ GeV (left panel) and 10 GeV (right panel) as a function of the quark
transverse momentum $|\kperp|$ for three different values of $x$ = 0.05, 0.1, and 0.2.
We select these three values of $x$ in order to show the TMD PDF
in the region covered by the experimental measurements (see Fig.~\ref{f:Data_Coverage}).
The error bands correspond to the 68$\%$ CL. They reflect the uncertainty on the fitting parameters
of Eq.~\eqref{e:pif1NP} that are determined by propagating the error in the
collinear PDFs of both the pion and the proton. However, since we used
only the central member of the set of collinear PDFs to perform the fit, the integral in $\kperp$
of all TMD replicas is fixed, \textit{i.e.}, their value at $\bT=0$ is the same.
As a consequence, the uncertainty related to collinear PDFs is only partially accounted in the plots.

We notice that in both the left and right panels of Fig.~\ref{f:tmds} the TMD PDF at $x$ = 0.05 shows the largest
error band, particularly at small values of $|\kperp|$. This kinematic region is at the boundary
of the phase space covered by the considered experiments. Future data from
the COMPASS Collaboration are expected to play an important role in improving this picture.

The mean squared transverse momentum of quarks in the pion
at $Q= 1$ GeV and $x = 0.1$ corresponds to the parameter $N_{1 \pi}$ of our fit
and turns out to be $\langle k_\perp^2\rangle  = 0.47 \pm 0.12$, somewhat
larger than the corresponding quantity for the proton (see Fig.~16 or
Ref.~\cite{Bacchetta:2022awv}). Therefore, our analysis indicates that the
TMD of quarks in the pion is
wider than that in the proton.

\subsection{Predictions at COMPASS kinematics}
\label{ss:PredCOMPASS}

The COMPASS Collaboration has recently released data for (un)polarized azimuthal asymmetries in the (polarized) pion--induced Drell--Yan processes~\cite{COMPASS:2017jbv,10.21468/SciPostPhysProc.8.028}, and will probably release in the near future also data for the unpolarized cross section.

\begin{figure}[h]
	\centering
	\includegraphics[width=0.9\textwidth]{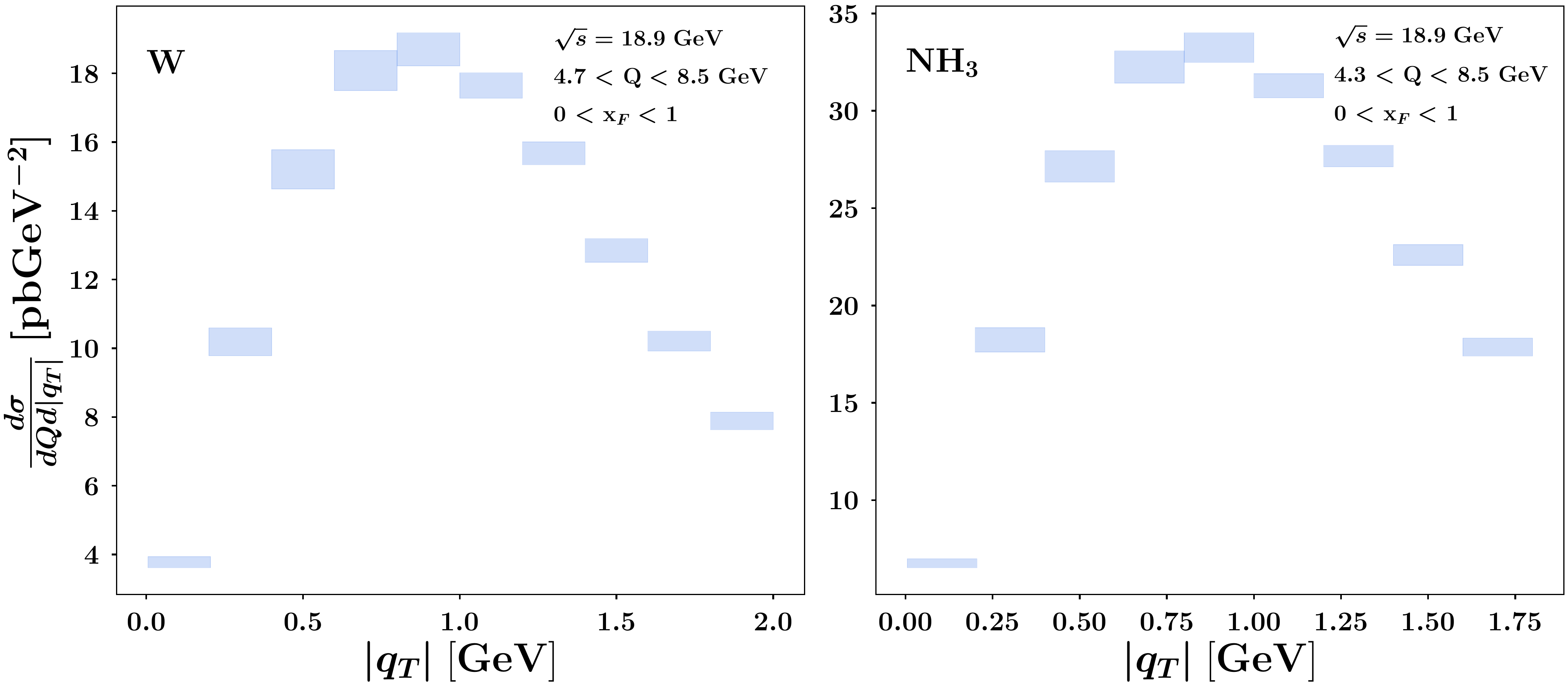}
	\caption{Theoretical predictions based on our fit for the unpolarized pion--nucleus DY cross section as function of the virtual vector boson transverse momentum $|\qT|$. Left panel for tungsten ($W$) nucleus, right panel for ammonia molecule (NH$_3$). Uncertainty bands correspond to 68\% CL.}
	\label{f:COMPASS_pred}
\end{figure}

In Fig.~\ref{f:COMPASS_pred}, we show theoretical predictions based on our fit for the unpolarized pion--nucleus DY cross section as function of the virtual vector boson transverse momentum $|\qT|$. The left panel refers to the tungsten ($W$) nucleus, while the right panel to the ammonia molecule (NH$_3$). The kinematics is the same of Ref.~\cite{10.21468/SciPostPhysProc.8.028} and is similar to the one covered in Fig.~\ref{f:Data_Coverage} by the experimental data analyzed in our fit.
The uncertainty bands at 68\% CL are evidently large, indicating that the available information we have on the internal structure of the pion is not sufficient to make accurate predictions.
Hopefully, the upcoming COMPASS data for the pion--nucleus DY process will help to better constrain TMD PDFs in the  pion, in particular shedding light on the normalization issue between theory and data.

\section{Conclusions and outlook}
\label{s:conclusions}

In this paper, we have presented an extraction of the unpolarized quark TMD PDF in the
pion, based for the first time on the analysis of the whole set of available
data for the production of DY lepton pairs
in $\pi^-$-nucleus collisions from the E615~\cite{Stirling:1993gc} and
E537~\cite{Anassontzis:1987hk} experiments. Our data set includes 138 data
points that can reasonably be described in the TMD formalism, in terms of TMD PDFs of the proton and of the pion.
The information about the proton is fixed
according to a recent extraction from the MAP
collaboration~\cite{Bacchetta:2022awv}, based on the analysis of a much larger
data set.
We modeled the pion TMD PDF as a simple Gaussian with an $x$-dependent width,
described by three free parameters in total.

We obtained a fairly good description of the data, with a global $\chi^2$ per
data point of 1.55. We emphasize that about 60\% of this value
comes from fully correlated normalization errors.

The TMD PDFs obtained by our fit, illustrated in Fig.~\ref{f:tmds}, are naturally affected by
larger error bands compared to proton TMD PDFs, even using a simple and rigid
functional form.

Our results can be used also to make predictions for future measurements. For
instance, we provided predictions for the unpolarized cross section in pion--nucleus DY collisions at the COMPASS kinematics.

\begin{acknowledgments}
  We thank Barbara Pasquini for discussions related to the structure of the pion
  and Riccardo Longo for
  discussions concerning COMPASS predictions.
  This work is supported by the the European Union's Horizon 2020 programme
  under grant agreement No. 824093 (STRONG2020).
  CB\ is supported by the DOE contract DE-AC02-06CH11357.
\end{acknowledgments}
\bibliography{PionMAPTMD22.bib}
\end{document}

%% file: settings.tex
\usepackage[paperwidth=215.9mm,paperheight=279.4mm,centering,hmargin=2cm,vmargin=2.5cm]{geometry}

\usepackage[tbtags]{amsmath}  
\usepackage{mathrsfs}
\usepackage{tabularx}
\usepackage{amssymb}          
\usepackage{bm}               
\usepackage{graphicx}         
\usepackage[export]{adjustbox}
\usepackage{hhline,multirow}  
\usepackage{makecell}  
\usepackage{dcolumn}          
\usepackage{slashed}
\usepackage{datetime}
\usepackage{color}
\usepackage[dvipsnames]{xcolor}
\usepackage{slashed}
\usepackage{subfloat}
\usepackage{float}
\usepackage{amscd}            
\usepackage{epsfig}
\usepackage{dcolumn}          
\usepackage[dvipsnames]{xcolor}
\usepackage[normalem]{ulem}
\usepackage{mathtools}
\usepackage{esint}
\usepackage{comment}
\usepackage[utf8]{inputenc}

\RequirePackage{color}
\RequirePackage[colorlinks=true
,urlcolor=black
,anchorcolor=black
,citecolor=black
,filecolor=black
,linkcolor=black
,menucolor=black
,pagecolor=black
,linktocpage=black
,pdfproducer=medialab
,pdfa=true
]{hyperref}

\newcommand{\nocontentsline}[3]{}
\newcommand{\tocless}[2]{\bgroup\let\addcontentsline=\nocontentsline#1{#2}\egroup}

\newcommand{\kperp}{\boldsymbol{k}_\perp}

\newcommand{\bT}{\boldsymbol{b}_T}

\newcommand{\qT}{\bm{q}_{T}}